# Structural stability and defect-tolerance of ionic spinel semiconductor for high-efficiency solar cells

Hanzhen Liang,[⊥] Huiwen Xiang,[⊥] Rui Zhu, Chengyan Liu[*], Yu Jia

Key Laboratory for Special Functional Materials of Ministry of Education, School of Materials Science and Engineering, Henan University, Kaifeng, Henan 475001, China

E-mail: cyliu@henu.edu.cn

## Abstract

The incompatibility between defect-tolerance and structural stability is a severe issue hindering the wide application of high-efficiency solar cells. Usually, covalent/polar semiconductors with prototype of Si/CdTe crystals exhibit great structural stability due to their compact tetrahedral structure, yet they present extremely poor defect-tolerance arising from the similar electronegativity of their component elements. On the contrary, ionic semiconductors, such as perovskite series, always exhibit favorable electronic properties of intrinsic defects caused by the great disparity of electronegativity between anions and cations, but are structurally unstable because of the sparsely composed octahedral building blocks supported by large cations. Combining the stable framework of covalent semiconductors and benign defects of ionic compounds, we find that $HgX_2S_4$ (X=In, Sc and Y) spinel semiconductors possess both the merits. The tightly combined tetrahedral and octahedral building blocks ensures the structural stability, and the band edge of ionic characteristic, mainly dominated by Hg-6$s$/X-$ns(d)$ for conduction band minimum (CBM) and S-3$p$ orbitals for valence band maximum (VBM), makes $HgX_2S_4$ defect-tolerant. The prominent downward bending of CBM caused by spatially spreading Hg-6$s$/X-$ns(d)$ orbitals with a large principal quantum number ($n$) not only induces a suitable optical band gap which is often too large in ionic compounds, but also promotes the formation and transport of n-type carriers. This study presents that Hg-based chalcogenide spinels are promising candidates for high-efficiency solar cells, and suggests that adopting cations with delocalized orbitals under the framework of spinel crystal is an alternative way for synthesizing the stable and defect-tolerant photovoltaic materials.

# 1. Introduction

Defect-tolerance and structural stability are two essential prerequisites for the wide application of high-efficiency solar cells. However, these two merits are difficult to be compatible in generally recognized semiconductors. For instance, despite the compactly combined robust tetrahedron structures of covalent/polar semiconductors (Si/CdTe), their intrinsic defects exhibit deep-level defect states resulting in a substantial deterioration of photovoltaic performance.[1-3] In contrast, perovskite solar cells dominated by ionic characteristic show benign electronic properties of intrinsic defects,[4,5] but have very poor environmental durability because of the sparse connected octahedron structures supported by large cations.[6-8] Seemingly, a feasible way to possess these two advantages is to construct a tightly combined tetrahedral or octahedral structure held by ionic bonds. Unfortunately, such a structural feature with the prototype of ZnO usually presents an excessive optical band gap resulting from the large disparity of electronegativity between anions and cations.

Usually, structure and electronic properties depending on the covalent/ionic types of semiconductors produce great effects on the applications of solar cells either in a beneficial or detrimental way.[9-13] Understanding the origin of structural stability and benign electronic property of existing semiconductors is of great importance for designing the stable and defect-tolerant photovoltaic materials. Conventional semiconductors, such as Si, CdTe and GaAs compactly composed by covalent/polar tetrahedron blocks exhibit robust structural stability. Their valence band maximum (VBM) and conduction band minimum (CBM) are derived from bonding and anti-bonding coupling of atomic orbitals with approximate energy.[14,15] Electronic structure in such a characteristic determines that energy levels of intrinsic point defects or dislocations with either insufficient or redundant unpaired electrons will be almost restored to their atomic values, hence are deeply located in band gap.[2] Even worse, grain boundaries of some particular polar semiconductors, such as CdTe and $Cu_2ZnSn(SSe)_4$, always produce deep-level defect states originating from the bonding states of paired anions,[16-19] which increase the carrier nonradiative recombination rate significantly. Although most of the conventional semiconductors are treated by extrinsic elements to eliminate their detrimental defect states introduced by point defects, grain boundaries or dislocations before they are used as solar cells, their efficiencies are still stalemated due to the inevitable defect states.[16,20] Ionic crystal with the representative of metal oxide always presents a stable tetrahedral framework and shallow energy levels of intrinsic defects due to the large disparity of electronegativity between cation and oxygen.[2,26,27] However, this feature also results in a much larger band gap than the optimal single-junction value. In contrast, the recently discovered hybrid organic-inorganic perovskites not only exhibit a great defect-tolerance induced by ionic characteristic of band edges,[21-25] but also have an upward bending of VBM owing to the high-energy spatially spreading Pb-6$s$ orbital, which makes up the disadvantage of excessive band gap for ionic crystals.[26-28] However, their sparsely connected octahedral building blocks supported by large cations have a great rotational freedom,[29-31] and the supporting cations are easily lost from the surface because of their weak interaction with anions.[32,33] These lead to the structural instability of perovskites under the environment of heat,[34] humidity[22,35] and oxygen.[35] In short, although conventional covalent/polar, ionic and perovskite crystals have their individual advantages, none of them simultaneously meet all the essential conditions of suitable band gap, structural stability and defect-tolerance required by an ideal photovoltaic material.

In this work, based on the cognition of structural stability, defect-tolerance and band edge composition, we find that the spinel system with the compact structures of tetrahedral and octahedral building blocks, which are formed by anion clusters centered by heavy cations, is a promising candidate for high-efficiency solar cell. Taking spinels of Hg-based chalcogenide ($HgX_2S_4$, X = In, Sc and Y elements) as examples, which are screened from the high-throughput calculations recently,[36] the mixed tetrahedral and octahedral S clusters centered by Hg and X, respectively, ensures the structural stability. Their CBMs mainly derived from the spatially spreading Hg-6$s$/X-$ns$($d$) orbital bend downward prominently, not only leading to a suitable optical band gap that is usually too large for ionic crystals, but also facilitating the formation of n-type defects and fast transport of carriers. Meanwhile, the prominent ionic characteristic of $HgX_2S_4$ with great difference of components between CBM dominated by Hg-6$s$/X-$ns$($d$) and VBM dominated by S-3$p$ results in the great defect-tolerance.

In the second part of this article, we describe the crystal structure of spinel and detailed calculation methods. In the third part, we analyze the electronic structures of $HgX_2S_4$ systems and predict the shallow energy level of intrinsic defects by normalized orbital overlap (Noo).[37] The fourth part aims at phase diagram, formation energy of intrinsic and Cl doping defects as well as their transition energy levels. In the fifth part, we focus on calculating the concentration of carries involving degeneracy factor $g_q$ of defect states, Fermi energy levels and defect concentrations. The sixth part is dedicated to the establishment of photovoltaic device with electron- and hole-transporting materials (ETM and HTM) through band alignment. Finally, the seventh part concludes all the results and suggests the way forward to high-efficiency solar cells in spinel framework.

## 2. Spinel structure and detailed calculation methods

### 2.1. Structural characteristic of spinel

Spinel, as shown in Fig. 1, mixing the tetrahedral and octahedral building blocks in a single crystal, enhances the structural stability by restricting rotation and sliding degrees of freedom. Its high crystal symmetry (Fd$\bar{3}$m), despite lower than the highest $O_h$ symmetry of conventional three-dimensional semiconductors and latest perovskites, not only allows the strong optical absorption at band edge because of inversion symmetry, but also facilitates the collection and transport of carriers owing to the isotropic low index crystal planes. Divalent Hg and trivalent X (X = In, Sc and Y elements) are located at the centers of tetrahedral and octahedral clusters, respectively, coordinated with S. The large disparity of electronegativity between metals and S contributes to the formation of ionic characteristic which leads to a great defect-tolerance. A high-energy level of Hg-6$s$/X-$ns(d)$ orbital exhibits spatially spreading characteristic in periodic crystals, and its anti-bonding state acting as the component of CBM or VBM exhibits a large dispersion in reciprocal space, making up the disadvantage of excessive band gap of ionic crystals. Therefore, HgX$_2$S$_4$ spinel is a promising candidate for the high-efficiency solar cells with suitable or adjustable band gap and great defect-tolerance.

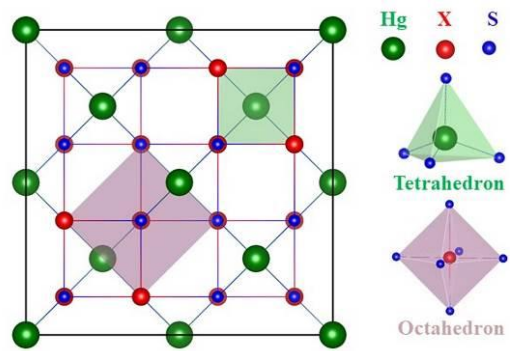

**Fig. 1** Spinel HgX$_2$S$_4$ (X = In, Sc and Y) with a high Fd$\bar{3}$m crystal symmetry composed by the tetrahedral and octahedral building blocks. Atoms colored green, red and blue represent Hg, X and S,

respectively. Cations (Hg and X) are in the centers of anion S coordinated tetrahedral and octahedral clusters, respectively.

**2.2. Detailed calculation methods**

The first-principles calculations based on the density functional theory are performed as implemented in the Vienna *ab initio* simulation package (VASP).[38,39] Structures with intrinsic defects are modeled using a 2×2×2 supercell. The projected-augmented-wave with Perdew-Burke-Ernzerhof (PBE) function is used in all the simulations. All of the pristine defect structures have been fully relaxed with Hellmann-Feynamn forces less than 0.01 eV/Å and electronic convergence less than $10^{-4}$ eV. The energy cutoff is up to 350 eV for the sufficient plane-wave expansion, and k points are sampled with a Γ-centered 5×5×5 mesh for the integration in Brillouin zone.[40] Based on the fully relaxed structures by the aforementioned PBE calculations, hybrid density functional method (HSE06) with one quarter portion of mixing coefficient is used to calculate the electronic structures and total energies of the concerned systems. Finally, the corrected band gaps by moving CBM to $GW_0$ results mentioned in previous study[36] are used to calculate electron concentration.

## 3. Electronic structure and normalized orbital overlap

**3.1. Electronic structure of HgX$_2$S$_4$ spinels**

Fully understanding the electronic structures of HgX$_2$S$_4$ spinel is the key to design defect-tolerant solar cells. Three types of HgX$_2$S$_4$ with different electronegativity of X (In: 1.78, Sc: 1.36 and Y: 1.22) have been studied. Generally, CBM states of most metal oxides/sulphides are derived from spatially spreading spherical *s* orbital of metals. Especially for the outer *ns(d)* orbital with a large principal quantum number *n*, CBM states exhibit more delocalized characteristics in reciprocal space. Therefore, electrons in CBM of metal oxides/sulphides have very small effective masses, and materials always present n-type characteristic due to the significant downward bending of CBM. In contrast, their VBM states are always dominated by O-2*p*/S-3*p* orbitals which are rather deep and localized in band structure. These make hole doping difficult and lead to a relatively large hole effective mass.

For HgIn$_2$S$_4$, its VBM states come from the localized S-3*p* orbital, as shown in Fig. 2a and d, and the states of Hg-5*d* and In-4*d* orbitals are slightly lower than the VBM. The downward bending of CBM is mainly caused by the very delocalized Hg-6s and In-5s orbitals. For HgSc$_2$S$_4$ (see Fig. 2b and e) and HgY$_2$S$_4$ (see Fig. 2c and f), their VBM states are dominated by S-3p orbital, which exhibits prominent ionic characteristics owing to the lower electronegativity of Sc and Y than that of In in HgIn$_2$S$_4$. Similar to HgIn$_2$S$_4$, CBM of HgSc$_2$S$_4$ and HgY$_2$S$_4$ also stems from the outmost orbitals of cations, such as Hg-6s and Sc-3d for HgSc$_2$S$_4$ and Hg-6s and Y-4d for HgY$_2$S$_4$. Besides the delocalized Hg-6s, Sc-3d and Y-4d also present spatially spreading characteristic for their relatively large principal quantum numbers, illustrated by the connected nd states shown in Fig. S2. Therefore, both the Hg-6s and X-nd orbitals in HgY$_2$S$_4$ result in the downward bending of CBM, together with the ionic characteristic of band edges, which promotes HgX$_2$S$_4$ to exhibit great defect-tolerance and suitable optical band gap. Meanwhile, HgX$_2$S$_4$ systems display a relatively high optical absorption coefficient (~10$^4$ cm$^{-1}$) within the visible light range as shown in Fig. S1 of Supporting Information (SI).

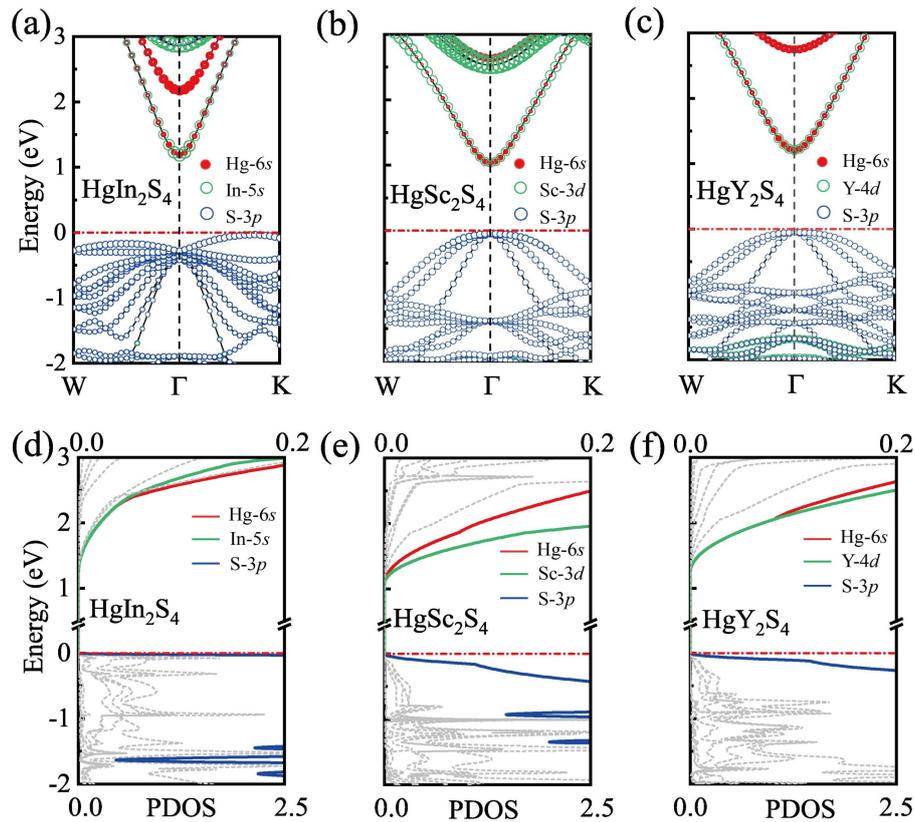

**Fig. 2** Band structure and projected density of states (PDOS) of HgIn$_2$S$_4$ ((a) and (d)), HgSc$_2$S$_4$ ((b) and (e)) and HgY$_2$S$_4$ ((c) and (f)). VBM of these three systems are mainly derived from S-3p orbital. CBM of HgIn$_2$S$_4$ is composed by the hybridization of Hg-6s and In-5s. Similarly, the hybridization of Hg-6s and Sc-3d contributes to the CBM of HgSc$_2$S$_4$, and Hg-6s and Y-4d to the CBM of HgY$_2$S$_4$. Red solid

circle, green hollow circle and blue hollow circle represent the components of Hg-6$s$, X-5$s$/3$d$/4$d$ and S-3$p$, respectively, whose size indicates their weights in band structures. The red dashed lines represent the Fermi levels.

**3.2. Normalized orbital overlap**

Noo as a function of the similarity of band edge composition is a great indicator to qualitatively evaluate the defect-tolerance of semiconductors.[37] The closer the Noo is to one, the more similar the orbital compositions of CBM and VBM are, which means the covalent property of crystals leading to deep levels of intrinsic defects. On the contrary, the closer the Noo is to zero, the greater the difference of band edge compositions is, suggesting a prominent ionic characteristic of crystals which exhibit a great defect-tolerance of intrinsic defects. The explicit Noo given by two manifolds of energy windows[37] is defined as

$$\text{Noo} = \langle \alpha | \beta \rangle \tag{1}$$

where

$$\langle \alpha | = c \begin{bmatrix} \rho_{v_1} & \rho_{v_2} & \cdots & \rho_{v_N} \end{bmatrix} \tag{2}$$

In eqn (2), $c$ is a normalization constant, and $\rho_{v_i}$ is the projected density of states of atomic orbital $v_i$. $i$ from 1 to $N$ with an interval of one discretizing $\rho_{v_i}$ facilitates the numerical integration of eqn (1).

In this study, two energy windows located at 1 eV above CBM and below VBM are selected for the calculations of $|\alpha\rangle$ and $|\beta\rangle$, respectively. The values of Noo for HgX$_2$S$_4$ systems are obtained and listed in Table 1. It can be seen that whether it is for the overall or individual orbital interaction between anions and cations, all the Noo values are less than 0.5 for these three HgX$_2$S$_4$ systems. This indicates that Hg-based chalcogenide spinel is expected to exhibit great defect-tolerance, which is consistent with the predictions of electronic structure analysis. In addition, it is seen that Noo value of Y-S (0.42) is larger than that of Sc-S (0.28), which seems abnormal for the larger electronegativity of Sc (1.36) than that of Y (1.22). It is caused by the stronger interaction between Sc-3$d$ and S-3$p$ than that of Y-4$d$ and S-3$p$ due to the same principal quantum number of the former, demonstrated by the shorter bond length of Sc-S (2.58 Å) than that of Y-S (2.74 Å). Therefore, the larger difference of Sc-3$d$ between VBM and CBM of HgSc$_2$S$_4$

than the situation of Y-4*d* in HgY$_2$S$_4$ leads to the smaller Noo value of Sc-S than that of Y-S.

**Table 1.** Noo between two manifolds of bands located at 1 eV above CBM and below VBM for spinel HgX$_2$S$_4$. All the Noo values are less than 0.5 for these three HgX$_2$S$_4$ systems, whether it is for the overall or individual orbital interaction between anions and cations.

| Spinel | Interaction | Noo |
|---|---|---|
|  | Total | 0.41 |
| HgIn$_2$S$_4$ | Hg-S | 0.49 |
|  | In-S | 0.49 |
|  | Total | 0.26 |
| HgSc$_2$S$_4$ | Hg-S | 0.39 |
|  | Sc-S | 0.28 |
|  | Total | 0.32 |
| HgY$_2$S$_4$ | Hg-S | 0.38 |
|  | Y-S | 0.42 |

## 4. Phase diagrams and defects properties

### 4.1 Stable phase diagrams

A phase diagram gives the conditions, e.g. elements partial pressures and temperature, at which distinct phases occur under thermodynamic equilibrium growth. The concerned spinel phases of HgX$_2$S$_4$ should satisfy

$$\mu_{Hg} + 2\mu_X + 4\mu_S = \Delta H_{HgX_2S_4} \tag{3}$$

where $\mu_i$ ($i$ = Hg, X and S) represents the chemical potential of element $i$ referred to its most stable pure elemental phase. In principle, the real thermodynamic stability of material is determined by Gibbs free energy depending on the internal energy, atmosphere pressure and temperature. For solids, the part of Gibbs free energy contributed by atmosphere pressure can almost be negligible. The contribution of vibrational and configurational entropy is usually very small for ordered crystals at room temperature. Therefore, the formation enthalpy $\Delta H_{HgX_2S_4}$ which is calculated with the only consideration of internal energy equals to -4.72 eV, -11.14 eV and

-12.10 eV for X = In, Sc and Y involved HgX$_2$S$_4$ systems, respectively.

Based on the chemical potential conditions described in the SI part 1, the stable phase diagrams as a function of the elemental chemical potential of HgX$_2$S$_4$ are obtained, as shown in Fig. 3. Since phase diagrams of the three HgX$_2$S$_4$ systems are restricted by avoiding the formation of pure elemental phases (S and Hg) and compounds of X$_2$S$_3$ and HgS, the regions of HgIn$_2$S$_4$, HgSc$_2$S$_4$ and HgY$_2$S$_4$ phase diagrams have a similar shape. It is known that Y has a lower electronegativity (1.22) than that of In (1.78) and Sc (1.36), thus Y$_2$S$_3$ ( $\Delta H_{Y_2S_3} = -11.64$ eV ) is the most easily formed secondary phase among X$_2$S$_3$ phases ( $\Delta H_{In_2S_3} = -10.60$ eV and $\Delta H_{Sc_2S_3} = -4.03$ eV ), resulting in the narrowest region of HgY$_2$S$_4$ phase diagram.

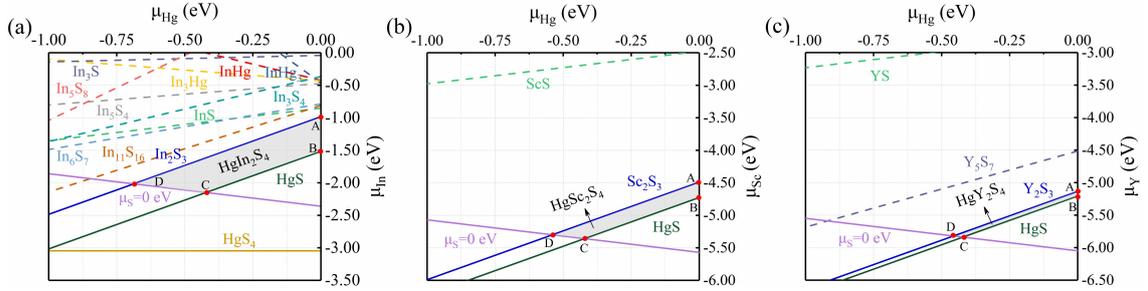

**Fig. 3** The calculated stable phase diagrams of (a) HgIn$_2$S$_4$, (b) HgSc$_2$S$_4$ and (c) HgY$_2$S$_4$ as a function of cation chemical potentials marked by gray area.

### 4.2. Defects properties

Formation energy directly reflects the formation ability of defect, which represents the energy cost of a certain number of atoms and electrons exchanging between host and atomic and electronic reservoirs.[41,42] Therefore, the formation energy $\Delta H_{\alpha,q}(\mu_i, E_F)$ of defect $\alpha$ in charge state $q$ can be defined as

$$\Delta H_{\alpha,q}(\mu_i, E_F) = \Delta E(\alpha, q) + \sum_i n_i \mu_i + qE_F \quad (4)$$

$$\Delta E(\alpha, q) = E(\alpha, q) - E(\text{host}) + \sum_i n_i \mu_i^0 + qE_{VBM} \quad (5)$$

where $E_F$ is the Fermi energy level referenced to the VBM of HgX$_2$S$_4$ host represented by $E_{VBM}$. $\mu_i$ is the chemical potential of atom $i$ relative to $\mu_i^0$ of the pure elemental phase of $i$. $E(\alpha, q)$ is the total energy of the relaxed supercell containing a defect $\alpha$ in charge state $q$. $E(\text{host})$ is the total energy of a defect-free supercell with the same size as the corresponding defect system. Defect transition energy level $\varepsilon_\alpha(q/q')$ can be viewed as the Fermi energy level where the formation energy $\Delta H_{\alpha,q}(\mu_i, E_F)$ of defect in charge state $q$ is equal to that of the same defect in charge state $q'$. Thus, the transition energy level is expressed as

$$\varepsilon_\alpha(q/q') = [\Delta E(\alpha,q) - \Delta E(\alpha,q')]/(q'-q) \qquad (6)$$

Formation energies of defects are calculated at the chemical potential conditions of A and C points which are representative on the diagonal of spinel HgX$_2$S$_4$ phase diagrams. It is seen that n-type defects of these systems have lower formation energies than that of p-type defects as Fermi energy level within the most range of band gap, as shown in Fig. 4a and b for HgIn$_2$S$_4$, 4c and d for HgSc$_2$S$_4$ and 4e and f for HgY$_2$S$_4$ which are corresponding to the conditions of A (Fig. 4a, c and e) and C (Fig. 4b, d and f) points in Fig. 3, respectively. This is consistent with the analysis of electronic structures that HgX$_2$S$_4$ (X= In, Sc and Y) systems have the low VBM dominated by S-3$p$ orbital and low CBM mainly induced by the spatially spreading Hg-6$s$/X-$ns(d)$ orbitals. As stated in doping limit rule,[43,44] a low VBM suppresses the formation of p-type defects as an electron at VB is difficult to be excited to form a hole, and a low CBM promotes the formation of n-type defects because it is easy to capture electrons. At A point, as shown in Fig. 4a, c and e, the cation-rich and anion-poor condition promotes the formation of n-type cation interstitial defects (Hg$_i$ and X$_i$) and anion vacancy defects (V$_S$), and suppresses the formation of p-type cation vacancy defects (V$_{Hg}$ and V$_X$). In contrast, at C point, as shown in Fig. 4b, d and f, the cation-poor and anion-rich condition facilitates the formation of p-type defects (V$_{Hg}$ and V$_X$) and restricts the formation of n-type defects (Hg$_i$, X$_i$ and V$_S$), while, too low band edges determine the n-type characteristic of Hg-based chalcogenide spinel systems in almost all regions of phase diagrams.

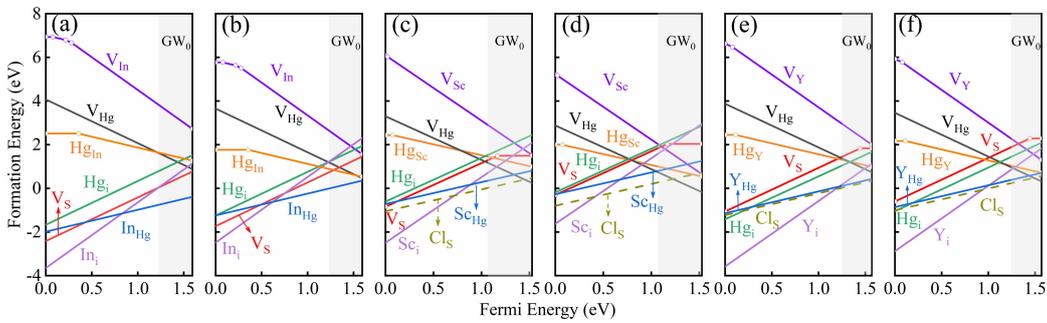

**Fig. 4** Defect formation energy as a function of Fermi energy level at chemical potential conditions of A and C points in Fig. 3. (a) and (b) for HgIn$_2$S$_4$, (c) and (d) for HgSc$_2$S$_4$ and (e) and (f) for HgY$_2$S$_4$

which are corresponding to the conditions of A (a, c and e) and C (b, d and f) points in Fig. 3, respectively. Cl doping forms $Cl_S^+$ defect in $HgSc_2S_4$ and $HgY_2S_4$ systems for the improvement of electron concentration, whose formation energies are exhibited in dashed lines. Band gap is corrected to $GW_0$ value by improving CBM to compensate the underestimation of band gap by PBE method.[45,46] The corrected parts are presented by gray regions. These corrected band gaps are used in the following calculations.

Transition energy levels of the intrinsic defects are shown in Fig. 5a-c for $HgIn_2S_4$, $HgSc_2S_4$ and $HgY_2S_4$, respectively. It is seen that almost all the transition energy levels of the dominant n-type defects are in the CB for these three systems, except for that of $V_S$ in $HgSc_2S_4$ and $HgY_2S_4$ which are still far away from the center of band gap under the corrected $GW_0$ results. Moreover, the formation energy of $V_S$ defects is relatively higher than that of the dominant n-type defects in $HgSc_2S_4$ ($Sc_{Hg}$ and $Sc_i$) and $HgY_2S_4$ ($Hg_i$ and $Y_i$), which are unlikely to act as the severe nonradiative recombination centers owing to their low concentrations. In contrast, almost all the p-type defects are suppressed due to the low VBM, most of which exhibit the relatively shallow transition energy levels. Overall, the characteristic of shallow transition energy levels of the dominant defects is consistent with the analysis of electronic structures and the Noo predictions in the section of 3.1 and 3.2.

The main reason for the high defect-tolerance of $HgX_2S_4$ materials are the ionic characteristic and the prominent downward bending of CBM caused by spatially spreading Hg-6$s$/X-$ns$($d$) orbitals. The ionic characteristic is conducive to the shallow defect levels near band edge, and the downward bending of CBM promotes the formation of n-type defects and makes most of the defect levels of n-type defects enter CB. The role of Hg-6$s$/X-$ns$($d$) orbitals in CBM of Hg-based spinel is very similar to the effect of spatially spreading Pb-6$s$ orbital in VBM of perovskite materials which not only facilitates the formation of p-type defect, but also promotes defect levels of p-type defects to locate at VB[4]. Therefore, CBM or VBM mainly constituted by spatially spreading $ns$($d$) orbital ($n$ is a relatively large principal quantum number) has a function of facilitating the defect-tolerance of n-type or p-type defects of solar cells.

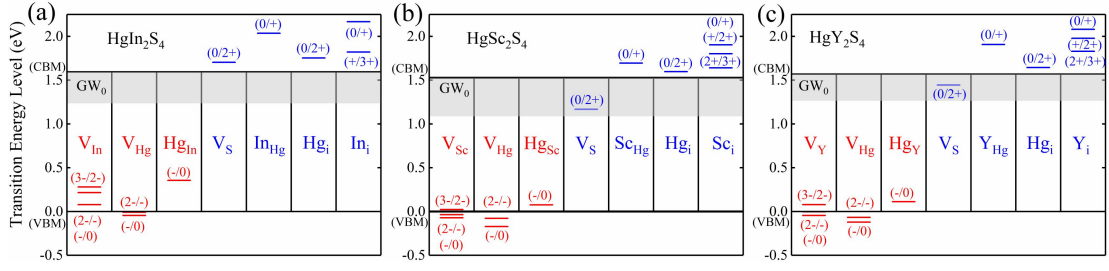

**Fig. 5** Transition energy levels of the intrinsic defects in band structures of (a) HgIn$_2$S$_4$, (b) HgSc$_2$S$_4$ and (c) HgY$_2$S$_4$. Red bars and blue bars represent the acceptor and donor transition energy levels, respectively, with the initial and final charge states labeled in parentheses. Band gaps by PBE calculations are corrected to the GW$_0$ values by adjusting the height of CBM shown in gray region.

## 5. The calculation of electron concentration

### 5.1. General theory of carrier concentration

Since carrier concentrations have great influence on the efficiency of solar cells, its basic theory is necessary to be introduced. For nondegenerate semiconductors, the concentrations of electrons ($n_0$) and holes ($p_0$) expressed in Boltzmann statistics are given by[47]

$$n_0 = N_C \exp\left(-\frac{E_C - E_F}{k_B T}\right), \qquad p_0 = N_V \exp\left(-\frac{E_F - E_V}{k_B T}\right) \qquad (7)$$

where $N_C$ and $N_V$ are the effective density of states of CBM and VBM, respectively, given by

$$N_C = \frac{2(2\pi m_n^* k_B T)^{\frac{3}{2}}}{h^3}, \qquad N_V = \frac{2(2\pi m_p^* k_B T)^{\frac{3}{2}}}{h^3} \qquad (8)$$

where $E_C$ and $E_V$ are the energies of CBM and VBM, which are set to $E_C = E_g$ and $E_V = 0$, where $E_g$ is the band gap of system. $k_B$ is Boltzmann constant, and $T$ is the temperature selected as 300 K. $m_n^*$ and $m_p^*$ are the effective mass of electron and hole at band edge, respectively. For degenerate semiconductors, Fermi energy level is close to band edges or enters CB or VB, then the general carrier concentration has to be expressed by Fermi-Dirac statistics which is used to describe the occupation of more electrons/holes in CB/VB. As Fermi energy level within band gap is several $k_B T$ away from band edge, Fermi-Dirac statistics degenerate to Boltzmann statistics. Here, the general Fermi-Dirac statistics are adopted to investigate the carrier concentrations of HgX$_2$S$_4$ spinel systems, which are given by[48,49]

$$n_0 = N_C \frac{2}{\sqrt{\pi}} (k_B T)^{-3/2} \int_{E_C = E_g}^{+\infty} \frac{(E - E_C)^{1/2}}{1 + \exp\left(\frac{E - E_F}{k_B T}\right)} dE,$$

$$p_0 = N_V \frac{2}{\sqrt{\pi}} (k_B T)^{-3/2} \int_{-\infty}^{E_V = 0} \frac{(E_V - E)^{1/2}}{1 + \exp\left(\frac{E_F - E}{k_B T}\right)} dE \qquad (9)$$

Considering the concentrations of positive charge $N_D^+$ and negative charge $N_A^-$ induced by donor ionization and acceptor ionization, respectively, the charge neutrality condition can be written as

$$p_0 + N_D^+ = n_0 + N_A^- \qquad (10)$$

Taking HgIn$_2$S$_4$ as an example, $N_D^+$ and $N_A^-$ are expressed as follows

$$N_D^+ = n_{\text{In}_{\text{Hg}}}^+ + 2n_{\text{Hg}_i}^{2+} + n_{\text{In}_i}^+ + 3n_{\text{In}_i}^{3+} + 2n_{V_S}^{2+}, \quad N_A^- = n_{V_{\text{In}}}^- + 2n_{V_{\text{In}}}^{2-} + 3n_{V_{\text{In}}}^{3-} + 2n_{V_{\text{Hg}}}^{2-} + n_{\text{Hg}_{\text{In}}}^- \qquad (11)$$

where $n_\alpha^q$ is the equilibrium concentration of an intrinsic defect $\alpha$ in charge state $q$. Combining with eqn (11), self-consistently solving eqn (10) can derive the Fermi energy level as a function of elemental chemical potentials. Based on the corresponding relation between Fermi energy level and elemental chemical potentials, the concentration of defect is obtained simultaneously.

**5.2. Degeneracy factor of intrinsic defects**

The degeneracy factor $(g_q)$ of a charged defect level determines the number of micro-states of electrons or holes on it, although usually the $g_q$ does not change the order of magnitude of defects and carrier concentrations. For an intrinsic defect $\alpha$ in charge state $q$ concerned in eqn (10) and (11), its equilibrium concentration is given by[50,51]

$$n_\alpha^q = N_{\text{sites}} g_q \exp\left(\frac{-\Delta H_{\alpha,q}(\mu_i, E_F)}{k_B T}\right) \qquad (12)$$

where $N_{\text{sites}}$ is the number of equivalent atomic sites for the formation of concerned defect per unit volume. $\Delta H_{\alpha,q}(\mu_i, E_F)$ is the formation energy described in eqn (4). For the counting of $g_q$, crystal field splitting of degenerate energy levels and electrons preferentially occupying the different orbitals of degenerate energy level in a spin-parallel way are considered under the Pauli exclusion principle. Taking $In_i$ defect as an example, triply degenerate $5p$ orbital of the interstitial In atom feeling a $D_{3d}$ crystal field splits into a low energy level of $5p_z$ singlet state and high energy level of $5p_x/5p_y$ doublet states as shown in Fig. 6a and b. Since a single defect level can hold two electrons with opposite spin at most, the neutral charge state of $In_i$ defect which has three electrons arranged in $5s$ and $5p_z$ orbitals will present $g_0 = 2$. After ionization, $g_+ = 1$ because of the only fully occupied lowest $5s$ orbital. Further ionization, we can get $g_{2+} = 2$ and $g_{3+} = 1$, as shown by the schematic diagrams in Fig. 6b. $g_q$ and the arrangement of electron on defect energy levels of other defects are listed in Table S1 of SI.

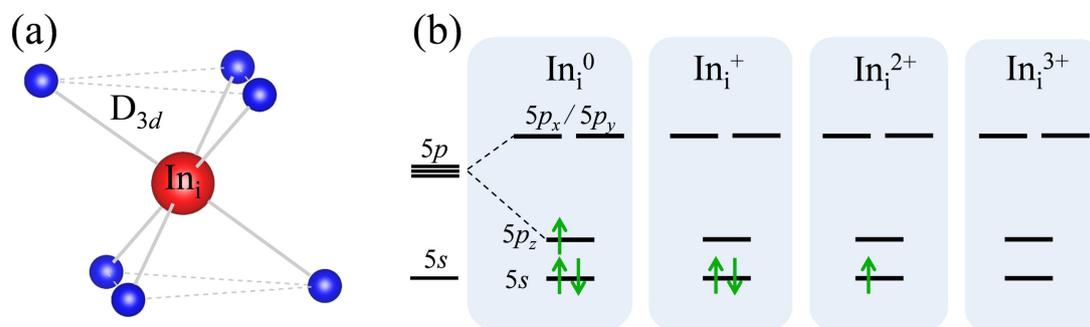

**Fig. 6** Atomic configuration and $5p$ orbital split of $In_i$ defect in $D_{3d}$ crystal field. (a) Atomic configuration of $In_i$ defect in $D_{3d}$ crystal field. Atoms colored blue and red represent S and In, respectively. (b) Triply degenerate $5p$ orbital splits into a low energy level of $5p_z$ singlet state and high energy level of $5p_x/5p_y$ doublet states, on which electrons of $In_i$ defect with different charge states are arranged.

### 5.3. Carrier concentration

Fermi energy level, as shown in Fig. 7a, is obtained by self-consistently solving the charge neutrality eqn (10) in which the defect concentration is expressed by eqn (12). It can be seen that intrinsic Fermi energy levels of the three $HgX_2S_4$ systems are all above the middle value of their band gaps (~ 0.75 eV) as shown in solid lines of Fig. 7a, indicating n-type characteristics of these systems. This is consistent with the analysis of electronic structures that the low VBM and CBM derived from low S-$3p$ orbital and greatly delocalized Hg-$6s$/X-$ns(d)$ orbitals, respectively, suppresses the formation of p-type defects and facilitates the formation of n-type defects based on doping limit rule.[43,44] Further, the Fermi energy level of $HgIn_2S_4$ is much higher than that of

HgSc$_2$S$_4$ and HgY$_2$S$_4$ owing to the lower formation energy of n-type In$_{Hg}^+$ defect than that of Sc$_{Hg}^+$ and Y$_{Hg}^+$ in the vicinity of CBM as shown in Fig. 4. The corresponding calculations of defect concentrations also confirm that the concentration of In$_{Hg}^+$ defect in HgIn$_2$S$_4$ is several orders of magnitude higher than that of Sc$_{Hg}^+$ in HgSc$_2$S$_4$ and Y$_{Hg}^+$ in HgY$_2$S$_4$, as shown in Fig. 8a-c.

Based on the understanding of defect concentration and Fermi energy level, the electron concentration of HgIn$_2$S$_4$ system should be higher than that of HgSc$_2$S$_4$ and HgY$_2$S$_4$, as confirmed by Fig. 7b. Generally, the optimal carrier concentration of absorbers is within $10^{15}$~$10^{18}$ cm$^{-3}$.[52,53] For HgIn$_2$S$_4$, the chemical potential condition of C point (both the In and Hg are poor ($\mu_{In}$=-2.15 eV and $\mu_{Hg}$=-0.42 eV)) suppresses the formation of n-type In$_{Hg}^+$ defect, which can lower the electron concentration into the optimal range marked by gradient color, shown by red solid line in Fig. 7b. While for HgY$_2$S$_4$, it has an appropriate electron concentration in the chemical potential range far away from the C point where both In and Hg are poor ($\mu_Y$=-5.84 eV and $\mu_{Hg}$=-0.42 eV), marked by green solid line in Fig. 7b. The synthesizing condition for the optimal electron concentration of HgY$_2$S$_4$ can be extended by an appropriate n-type doping. However, for HgSc$_2$S$_4$, it has the lowest electron concentration far away from the optimal carrier concentration in entire chemical potential range, shown by blue solid line of Fig. 7b.

Further improving the carrier concentration of HgSc$_2$S$_4$ and HgY$_2$S$_4$ systems can be realized by extrinsic n-type doping. Commonly, states of CBM and VBM of ionic crystals are derived from valence electron orbitals of cations and anions, respectively. Doping on the sites of cations or anions will more or less affect the position of CBM or VBM, the extent of which depends on topography of band edges. Since the CBM states of HgX$_2$S$_4$ are originated from spatially spreading Hg-6$s$/X-$ns$($d$) orbitals, doping on cation will lead to a large fluctuation of CBM which may introduce deep-level defect states and thus cause carrier nonradiative recombination. While their VBM states are derived from the relatively localized S-3$p$ orbital, and doping on anion will have less fluctuation of VBM. Therefore, Cl, which has one electron more than S, is an optimal dopant to improve electron concentration in HgSc$_2$S$_4$ and HgY$_2$S$_4$ systems by forming the n-type Cl$_S^+$ defect. Cl$_S$ defect of these two systems has the lowest formation energy than that of their intrinsic defects in the vicinity of CBM, illustrated by dashed lines in Fig. 4c-f. Apparently, both their Fermi energy levels and Cl$_S^+$ concentration are greatly improved compared with their intrinsic systems, shown by dashed lines in Fig. 7a and Fig. 8b and c, respectively. Finally, electron concentration of HgY$_2$S$_4$ is successfully adjusted to the optimal carrier concentration in the entire chemical potential range from A to C point, and HgSc$_2$S$_4$ also presents an optimal carrier concentration near A point, illustrated by green and blue dashed lines in Fig. 7b, respectively.

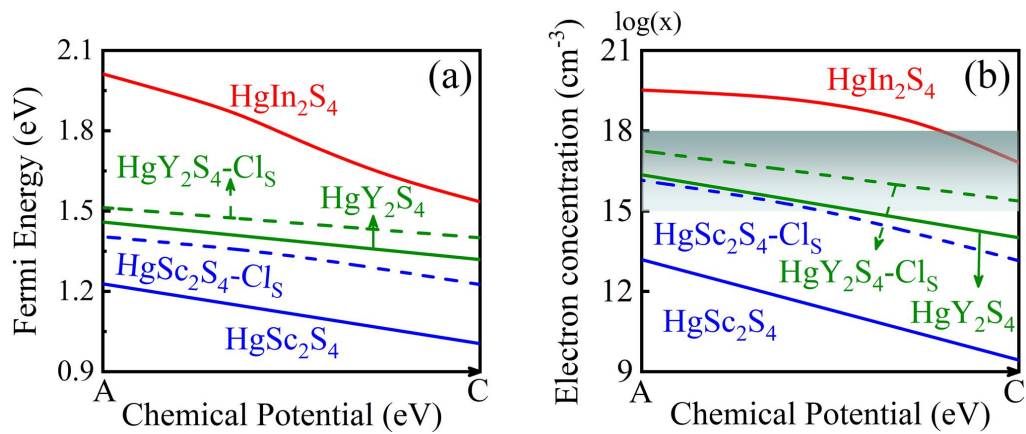

**Fig. 7** Fermi energy level and electron concentration of HgX$_2$S$_4$ as a function of elemental chemical potential from A to C points in Fig. 3. (a) Fermi energy level solved self-consistently from charge neutrality condition. (b) Electron concentrations of HgX$_2$S$_4$ systems as well as Cl-doped systems. HgIn$_2$S$_4$ has the highest Fermi energy and electron concentration than those of HgY$_2$S$_4$ and HgSc$_2$S$_4$. Solid lines represent the intrinsic characteristic of HgX$_2$S$_4$ systems. Green and blue dashed lines represent the properties of Cl-doped HgY$_2$S$_4$ and HgSc$_2$S$_4$ systems which raise the Fermi energy levels and electron concentrations.

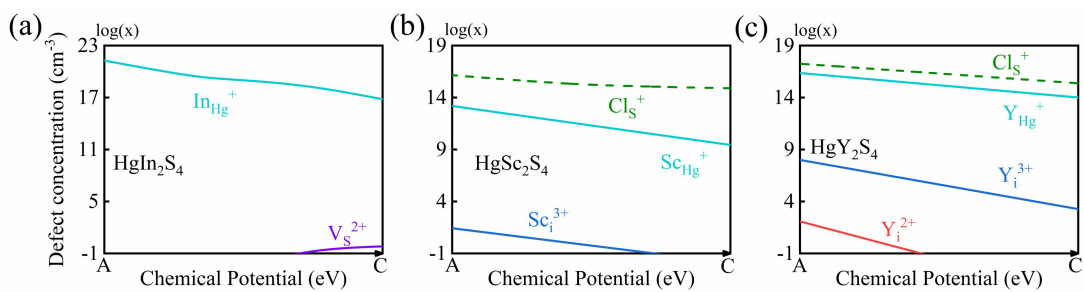

**Fig. 8** Defect concentrations as a function of elemental chemical potential for (a) HgIn$_2$S$_4$, (b) HgSc$_2$S$_4$ and (c) HgY$_2$S$_4$. In$_{Hg}^+$ has the highest concentration than that of other defects in HgX$_2$S$_4$ systems. Cl$_S^+$ is the dominant defect in HgSc$_2$S$_4$ and HgY$_2$S$_4$ systems, owing to its lowest formation energy.

## 6. Design of HgX$_2$S$_4$ photovoltaic device

ETM and HTM are very important to the design of photovoltaic devices, which directly affect the extraction rate of carriers (short-circuit current $J_{sc}$) and the value of open-circuit voltage ($V_{oc}$). Although low CBM of ETM and high VBM of HTM can promote the extraction of electrons and holes, too low CBM of ETM or too high VBM of HTM in devices usually cause a great loss of $V_{oc}$, which is not conducive to maximizing the photovoltaic efficiency. Generally, about 0.2 ~ 0.4 eV difference between band edge of absorber and CBM of ETM or VBM of HTM is a trade-off choice for both the large $J_{sc}$ and high $V_{oc}$.[54,55] Therefore, band edges of the selected ETM and HTM should be as close to the optimal values as possible.

Taking VBM of $HgIn_2S_4$ as the reference zero point, its CBM is at the gap value (1.60 eV) corrected by $GW_0$ calculation. Since the absorber exhibits n-type characteristic, electrons as majority carrier are easier to be extracted than that of the minority carrier of holes. Therefore, $TiO_2$ as ETM, the CBM of which is slightly higher than that of $HgIn_2S_4$, not only guarantees enough electrons to be extracted, but also does not lose the $V_{oc}$ on ETM side as shown in Fig. 9. Since holes are minority carriers, their extracting rate seriously affects the efficiency of solar cells. Therefore, HTM should have an appropriately higher VBM than that of absorber. Band alignment shown in Fig. 9 indicates that NiO whose VBM is 0.36 eV higher than that of absorber is very suitable as a HTM. Due to the approximate band edge positions, $HgIn_2S_4$ and $HgSc_2S_4$ can adopt the same ETM and HTM. The structures of $TiO_2/Hg(In,Sc)_2S_4/NiO$ which possess the optimized $J_{sc}$ and $V_{oc}$ are suggested as the design of high-efficiency solar cells. The same principle is applied to select ETM and HTM for $HgY_2S_4$ which had better have lower VBM and CBM than that for $Hg(In,Sc)_2S_4$ due to the lower VBM and CBM of $HgY_2S_4$. Therefore, $SnO_2$ and F8 with suitably aligned CBM and VBM, respectively, can be considered as the ETM and HTM for $HgY_2S_4$.

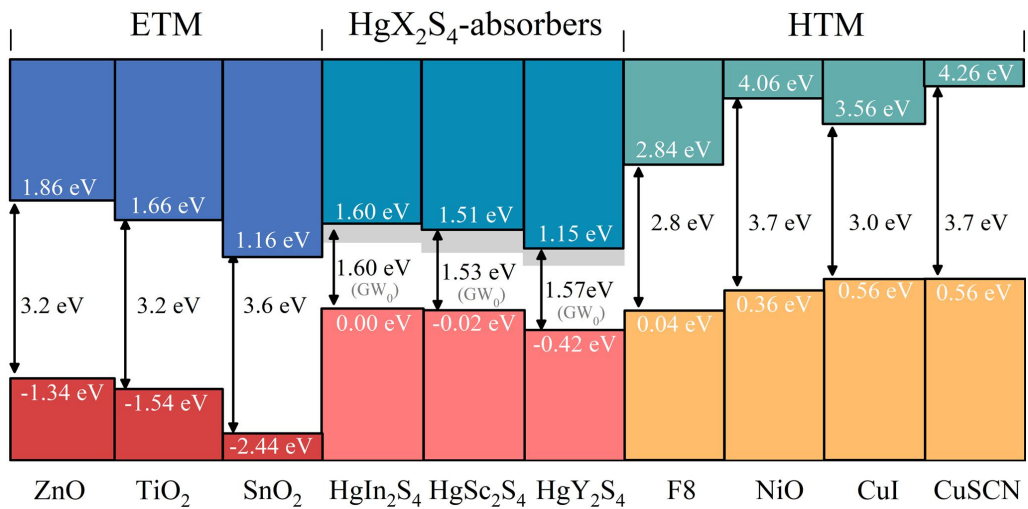

**Fig. 9** Comparison of band edge energetics between $HgX_2S_4$, ETM: ZnO, $TiO_2$ and $SnO_2$ and HTM: F8, NiO, CuI and CuSCN. Band edge positions of ETM and HTM are collected from Refs. 56-58. Band

alignments between HgX$_2$S$_4$ and ETM/HTM are calculated by core level alignment of anions without considering the connecting effect of two grains.

## 7. Conclusions

This work presents the first detailed study of HgX$_2$S$_4$ (X=In, Sc and Y) spinel crystals as the potential high-efficiency solar cells. Their structure feature of tightly combined tetrahedral and octahedral building blocks ensures the structural stability in working environment. Meanwhile, the great disparity of electronegativity between cations and anions makes the systems exhibit ionic characteristics, efficiently avoiding the formation of deep-level defect states. Moreover, the spatially spreading Hg-6$s$/X-$ns(d)$ spherical orbital leading to a prominent downward bending of CBM not only induces a suitable optical gap value which is always too large in conventional ionic crystals, but also promotes the fast transport of carriers. The carrier concentration can be further improved to the optimal range through the extrinsic Cl doping. The comparison of band alignments suggests that TiO$_2$/Hg(In,Sc)$_2$S$_4$/NiO and SnO$_2$/HgY$_2$S$_4$/F8 are reasonable device designs for high-efficiency solar cells. This study systematically presents the merits of Hg-based chalcogenide spinel crystals as high-efficiency solar cells in terms of structures, defects, carriers and design of devices, and suggests that under the spinel structural framework, replacing Hg by environment friendly cations with the similarly delocalized outer orbitals is an alternative way for synthesizing the environmentally durable and defect-tolerant photovoltaic materials.

**Author Contributions**

$^\perp$H.L. and H.X. contributed equally to this work.

**Notes**

The authors declare no competing financial interest.

**Acknowledgments**

This work was partly supported by the National Nature Science Foundation of China (No. 12004100, 12074099), and partly by Key Scientific Research Foundation for Universities of Henan Province (21A140006).